# Optimization and Control Technologies for Renewable-Dominated Hydrogen-Blended Integrated Gas-Electricity System: A Review

Wenxin Liu, Jiakun Fang, Senior Member, IEEE, Shichang Cui, Member, IEEE, Iskandar Abdullaev, Suyang Zhou, Xiaomeng Ai, Member, IEEE, Jinyu Wen, Member, IEEE

*Abstract*—**The growing coupling among electricity, gas, and hydrogen systems is driven by green hydrogen blending into existing natural gas pipelines, paving the way toward a renewable-dominated energy future. However, the integration poses significant challenges, particularly ensuring efficient and safe operation under varying hydrogen penetration and infrastructure adaptability. This paper reviews progress in optimization and control technologies for hydrogen-blended integrated gas-electricity system. First, key technologies and international demonstration projects are introduced to provide an overview of current developments. Besides, advances in gas-electricity system integration, including modeling, scheduling, planning and market design, are reviewed respectively. Then, the potential for cross-system fault propagation is highlighted, and practical methods for safety analysis and control are proposed. Finally, several possible research directions are introduced, aiming to ensure efficient renewable integration and reliable operation.**

## I. INTRODUCTION

As an efficient secondary energy carrier, hydrogen demonstrates significant potential for integrating renewable energy across multiple energy sectors, including electricity, heating, transportation, and industry [1]. However, constructing dedicated hydrogen pipelines involves high capital costs and lengthy development timelines, and blending hydrogen into the existing natural gas network emerges as a cost-effective alternative [2]. As reported by ENTSOG, the existing European natural gas pipeline network extends over 205,090 km [3] and possesses a transmission capacity of up to 24,226 GWh per day [4]. Under the future plan with a 20% hydrogen blending ratio

[5], the setup can accommodate up to 40-70.8 GW electrolyser [6], absorb over 200 TWh/a of renewable electricity, and achieve a reduction of 6%-7% in greenhouse gas emissions [7]. To maximize the utility of hydrogen-blended gas infrastructure, bidirectional interactions through energy conversion units such as hydrogen-blended gas turbines and power-to-hydrogen are encouraged, serving as flexible resources [8]. The integration gives rise to a hydrogen-blended gas-electricity system (HIGES), constituting an essential step toward a renewable-dominated energy system.

Countries around the world are actively advancing hydrogen blending technologies through policy support and demonstration projects. In Europe, the EU Hydrogen Strategy released in July 2020 identifies hydrogen blending as a transitional solution to leverage existing natural gas infrastructure [9]. Under the European Hydrogen Backbone Initiative, the strategy aims to expand the hydrogen transmission network to 58,000 kilometers by 2040, with approximately 60% consisting of repurposed natural gas pipelines [10]. Notable progress includes the HyDeploy project in UK, which successfully tested hydrogen blending ratios of up to 20%, demonstrating significant potential for carbon emission reduction without the need for major infrastructure modifications [11]. In Spain, the HyDeal Ambition project plans to supply green hydrogen produced by solar-powered electrolysis through the existing gas network at a target price of €1.5 per kilogram by 2030, approaching cost parity with fossil fuels [12]. In the United States, HyBlend supported by DOE is addressing key technical challenges, focusing on material compatibility, techno-economic analysis, and life cycle assessment [13]. In contrast, theoretical research and engineering practice of hydrogen-natural gas blending technology in China began relatively late. In 2024, the first 258 km high-pressure, long-distance natural gas pipeline was completed in Inner Mongolia, with a maximum transmission capacity of 1.2 billion m³/year and a hydrogen blending ratio of up to 10%. These efforts provide essential technical foundations for the development of future standards and the safe, efficient operation of blended gas systems.

Aforementioned demonstration projects have primarily addressed the technical feasibility and regulatory constraints of hydrogen-blended natural gas systems [8], [14]. For example, investigations have examined alterations in thermodynamic and transport properties [15], environmental impacts [8], and effects on component corrosion and operational reliability [16]. These studies also summarize potential risks associated with hydrogen blending. First, because the energy density of

This work was supported by the National Key R&D Program of China (No. 2022YFB2404000) and the National Natural Science Foundation of China (Grant No. 52177089).

W. X. Liu (ORCID: https://orcid.org/0009-0006-9934-5632), J. K. Fang (corresponding author, email: jfa@hust.edu.cn; ORCID: https://orcid.org/0000-0002-8208-5938), S. C. Cui, X. M. Ai, J. Y. Wen, are with the State Key Laboratory of Advanced Electromagnetic Technology, Huazhong University of Science and Technology, Wuhan, China.

S. Y. Zhou is with the School of Electrical Engineering, Southeast University, Nanjing, China.

I. Abdullaev is with New Uzbekistan University, Uzbekistan.

gaseous hydrogen is roughly one‑third that of natural gas, higher flow rates and operating pressures are required to maintain equivalent energy delivery, potentially exceeding existing infrastructure design limits [14]. Second, hydrogen can diffuse into material crystal lattices, inducing embrittlement and elevating the likelihood of pipeline and seal failures [17], [18]. Third, fluctuating hydrogen injection generates heterogeneous gas mixtures with variable combustion characteristics, which may damage end-use equipment and underscore the necessity of stringent gas quality control [19]. Although these analyses are confined to gas subsystems, they reveal additional operational constraints for safety and reliability [16] as well as risk propagation within integrated systems [20], thereby calling for insights from integrated system interactions.

Current research primarily focuses on two aspects, i.e., the simulation of the impacts of alternative gas injections on the gas system or IEGS and the optimal operation with hydrogen blending. For example, studies have explored changes in thermodynamic and transport properties after hydrogen injection [15], the environmental effects of varying hydrogen blending ratios [8], and the impact of hydrogen on component corrosion and operational reliability [16]. These simulation-driven studies have advanced understanding of hydrogen influence on gas behavior and infrastructure performance. However, they often fail to account for safety constraints in a controllable or actionable way. For instance, if hydrogen concentrations in specific network segments exceed regulatory thresholds, existing models provide limited quantitative guidance for mitigating risks within integrated energy system (IES) frameworks [21]., optimization-oriented research has emerged, focusing on operational decision-making for hydrogen injection in HIGES [21][22]. These studies shift attention from passive impact analysis to proactive system management strategies. Nevertheless, substantial limitations remain. Most existing models oversimplify physical processes, and lack integrated approaches to co-optimize electricity and gas networks. Safety constraints are often inadequately formulated or excluded, limiting the practical applicability of results.

Critical gaps persist in understanding the system-level interactions between gas and electricity networks, the optimization of coordinated operation under dynamic conditions and the long-term impacts of hydrogen integration on infrastructure resilience and reliability. Additionally, there is limited consensus on standardized modeling frameworks, security mechanisms, and control strategies that can address both normal operations and failure scenarios. This review aims to address critical challenges related to the operation and optimization of HIGES. It begins with an overview of the system architecture and highlights key technologies and application potential of hydrogen-blended natural gas systems. Subsequently, it examines modeling frameworks and operational strategies under normal conditions, with a focus on system planning, scheduling, and market mechanisms. The review then analyzes coordinated security mechanism and control strategy under potential failure scenarios. Finally, it identifies the major technical barriers and outlines promising directions for future research and development.

## II. OVERVIEW OF HYDROGEN BLENDING TECHNOLOGY AND APPLICATION POTENTIAL

In this section, the role of hydrogen blending in decarbonized electricity and natural gas systems with high renewable energy penetration is analyzed. First, the typical structure of a hydrogen-blended integrated gas-electricity system is introduced. Then, critical technologies enabling hydrogen-enriched natural gas systems and their prospective development pathways are discussed. Finally, the progress of existing demonstration projects worldwide is summarized.

### A. General Structure of hydrogen-blended integrated gas-electricity system

A general structure of a hydrogen-blended integrated gas-electricity system is illustrated in Fig. 1. This system integrates regional electricity, natural gas, and hydrogen subsystems into a tightly coupled multi-energy network. Energy conversion devices such as P2G units and gas power plants (GPP) serve as key interfaces, enabling flexible conversion between electricity and gas-based energy carriers. During off-peak periods, surplus renewable energy can be used for hydrogen production via electrolysis. The produced hydrogen may be stored in tanks and later blended with natural gas. Alternatively, high-purity hydrogen can be extracted using technologies such as pressure swing adsorption (PSA) or membrane separation to serve industrial demands including metal refining, ammonia synthesis, and fuel cell electric vehicles. This approach leverages existing gas infrastructure for hydrogen transport while ensuring compliance with safety and quality standards [8].

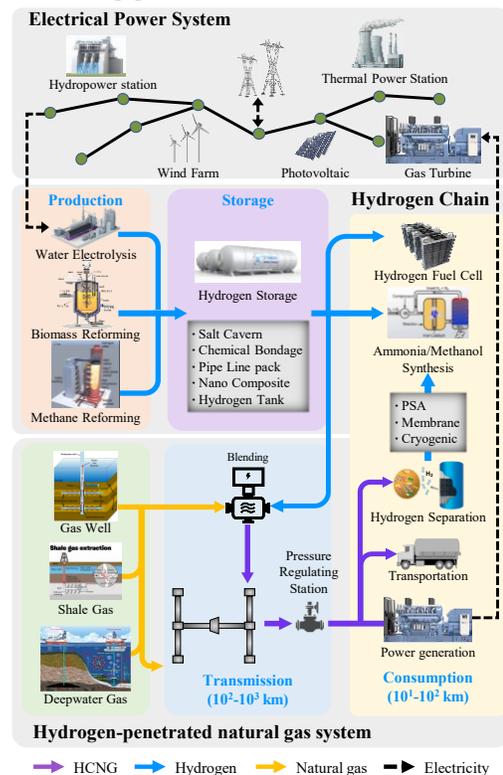

Fig. 1 General structure of hydrogen-blended integrated energy system

Hydrogen injection strategies significantly influence system performance. With multiple potential injection points, centralized or distributed injection may be employed [23]. Centralized injection, typically at source nodes, ensures a uniform gas composition throughout the network. Distributed injection, determined by the spatial availability of renewable resources, tends to affect only downstream sections, leading to uneven gas quality. For looped networks, the potential flow direction reversal after injection must also be considered [19], [24], [25]. Blending methods at injection points are equally critical and generally follow either a fixed ratio or fixed hydrogen flow rate. Most demonstration projects adopt dynamic fixed-ratio blending, while fluctuating blending ratios remain underexplored [14]. Key challenges include maintaining gas quality and ensuring that pressure and flow constraints are met across the network. Future studies should focus on evaluating the impacts of different hydrogen injection strategies to enable the safe and efficient delivery of hydrogen-blended natural gas.

### B. Critical technologies for hydrogen-blended natural gas system

Hydrogen-blended natural gas systems encompass four principal stages: production, storage, transmission, and utilization. The key technologies involved are summarized as follows.

1) Production: Surplus electricity can be converted into hydrogen through several electrolysis technologies, including alkaline electrolysis (AEL), proton exchange membrane electrolysis (PEM), solid oxide electrolysis (SOE), and anion exchange membrane electrolysis (AEM) [26], as summarized in TABLE I. Among them, AEL is the most mature and cost-effective, widely deployed in large-scale applications at the megawatt and gigawatt levels [27]. However, its performance is constrained by hydrogen crossover, especially at low loads, along with high overpotentials and slow ion transport in the electrolyte. These factors lead to limited load flexibility and reduced dynamic responsiveness [28]. PEM electrolysis improves upon AEL by offering higher current density, faster response, and broader load flexibility, making it particularly suitable for systems with high renewable energy penetration [28]. Nevertheless, its reliance on noble metal catalysts and selective membranes increases equipment costs relative to AEL [29]. SOE operates at temperatures of 700°C-1000°C, which enhances electrolysis efficiency by leveraging favorable thermodynamics and kinetics compared to low-temperature water electrolysis. Despite its potential, SOE requires additional heat sources and exhibits limited response speeds [30]. Currently, its application is confined to laboratory settings and demonstration projects. AEM combines the advantages of AEL and PEM, using an alkaline electrolyte to avoid costly noble metals while targeting high current densities. However, AEM remains at an early research stage and is not yet commercially viable [27].

Integration of multiple hydrogen production technologies is another promising research direction. For instance, [31] proposes optimal design and technology selection for electrolysis stations providing various grid services, such as demand response and renewable smoothing. In [28], a unified operational model for P2H systems in active distribution networks was proposed to improve flexibility and efficiency through optimized technology selection, site planning, and capacity sizing. Furthermore, [32] presents a hydrogen energy hub integrating water electrolysis and biomass gasification to produce stable green hydrogen, catering to diverse industrial demands. These integrated systems not only enhance the flexibility and resilience of power grids but also contribute to the efficient utilization of renewable energy sources, paving the way for a sustainable hydrogen economy.

TABLE I INTRODUCTION OF DIFFERENT P2H TECHNOLOGIES

| Parameters | | AEL | PEM | SOE | AEM |
|---|---|---|---|---|---|
| Charge carrier | | OH⁻ | H⁺ | O²⁻ | OH⁻ |
| Operating temperature[33] | °C | 70-90 | 50-80 | 700-1000[30] | 40-60 |
| Operating pressure[33] | bar | 1-30 | <70 | 1-25[34] | <35 |
| Load range[28] | % | 25-100 | 5-120[26] | 10-100 | 5-100[26] |
| System efficiency[35] | % | 50-78 | 50–83 | 50–89 | 57–59 |
| Voltage range[35] | V | 1.4–3.0 | 1.4–2.5 | 1.0-1.5 | 1.4–2.0 |
| Operating current density[35] | A/cm² | 0.2-0.8 | 1.0-2.0 | 0.3-1.0 | 0.2-2.0 |
| Cold start-up time[30] | | 1-2h | 5-10min | hours | 5-10min[36] |
| Warm start-up time[30] | | 1-5min | <10s | 15min | <10s |
| Lifetime, stack[35] | h | >60000 | 50000-80000 | ~20000 | 20000-60000 |
| TRL[27] | | 8-9 | 8-9 | 6-7 | 2-4 |
| Unit investment cost[30] | €/kW | 800-1500 | 1400-2100 | > 2000 | - |

2) Storage: The research on hydrogen storage methods is advancing with a focus on physical, material-based, and chemical storage techniques [37]. Compressed gas storage is well-established but limited by capacity and high operating pressures [38]. Cryogenic storage offers high-density storage but faces boil-off losses and energy-intensive cooling [39]. Solid-state storage materials like metal hydrides and adsorbents offer safe storage but are hindered by slow kinetics and high costs [40]. Chemical storage methods, such as liquid organic hydrogen carriers (LOHCs) and chemical hydrides, provide reversible storage but encounter issues with catalyst degradation and efficiency [41], [42]. Underground storage is suitable for large-scale applications but poses geological and environmental risks [43]. The challenge lies in enhancing storage efficiency, reducing costs, and developing infrastructure that can support large-scale hydrogen utilization, aligning with the growing demand for sustainable energy systems.

3) Transmission: The high-pressure transmission network consists of pipelines, compressors, and pressure regulation stations. Injecting hydrogen into transmission pipelines requires substantial investment in flow monitoring and metering stations, as harmonized and stable gas quality must be ensured [44]. Hydrogen blending also affects compressor performance curves and operating conditions, often requiring higher rotational speeds to reduce the risk of surge [45].

In comparison, the lower-pressure distribution network is generally considered less sensitive to hydrogen injection due to their lower operating pressure, higher capacity margins, and compatible pipeline materials [8]. However, higher hydrogen blending ratios can accelerate material degradation. As reported by NREL, when the hydrogen blend reaches 50%, the risk indices for corrosion and material defects in both distribution mains and service lines increase significantly [18]. This underscores the need for hydrogen-compatible upgrades to pipelines and related infrastructure. To quantify the impact of hydrogen embrittlement, Ref. [46] reviews the progress in quantitative research on hydrogen embrittlement mechanisms.

4) Utilization: Hydrogen-blended natural gas can be directly used for combustion in industrial heating, building applications, and residential settings, as well as for electricity generation in gas-fired power plants. Field demonstrations, including GRHYD in France and Sustainable Ameland in the Netherlands, have shown that hydrogen blends up to 20% by volume can be safely used without modifications to end-user appliances [27]. Gas turbines can operate with 1-5% hydrogen without technical changes and up to 10% with minor adjustments [8]. Siemens and General Electric have developed turbines capable of handling 30% and up to 100% hydrogen blends, respectively [47].

Another application involves extracting pure hydrogen from the blended gas for use in fuel cell vehicles or industrial synthesis processes such as ammonia and methanol production. Common separation technologies include pressure swing adsorption (PSA) and membrane-based systems. While PSA can achieve high-purity hydrogen (>99.9%), it requires substantial adsorbent volumes and incurs high capital and operational costs when processing low hydrogen concentrations [48]. Membrane separation is modular and compact, but palladium membranes face issues like poisoning and embrittlement [49], and polymer membranes typically require multi-stage setups with high energy consumption due to low selectivity [50]. Consequently, hydrogen separation remains a significant technical and economic challenge, especially for pipeline applications with low hydrogen content.

C. Global Demonstration Projects

Hydrogen blending has emerged as a key transitional decarbonization strategy, attracting significant policy support and technological innovation worldwide. TABLE I summarizes the demonstration projects and their distinguishing features. Despite extensive investigation, there is still no consensus on the maximum feasible hydrogen blending ratio; this "upper limit" must be determined through joint analysis of empirical data from field tests and results from theoretical modeling and simulation.

Regarding network configuration, ENTSOG asserts that it might not be feasible to gradually increase the hydrogen fraction in gas networks from 0 to 100%. Instead, once a certain "tipping point" is reached that makes a full transition to hydrogen more economical, a full transition to pure hydrogen becomes more economical [5]. The conclusion offers critical guidance for the planning and evolution of integrated hydrogen–gas–electricity systems.

TABLE I COMPARISON OF OPTIMAL PLANNING MODEL FOR THE ELECTRICITY–NATURAL GAS COUPLING SYSTEM

| Project | Country/Region | Timeline | Limit | Features |
|---|---|---|---|---|
| NATURLHY | Europe | 2004-2009 | 50% | The first demonstration project worldwide; Injected hydrogen into high-pressure transmission pipelines and delivered to end users via the distribution network. Systematically examined the effects of hydrogen blending, published equipment aging assessments, and refined standards and regulations for mixed gases [51]. |
| HyDeploy | UK | 2017-2023 | 20% | Injected hydrogen into the natural gas distribution network [11]. |
| H21 Leeds City Gate | UK | 2016-2019 | 100% | Explored the feasibility of converting the UK's urban gas distribution network to a 100% hydrogen network [52]. |
| HyBlend | USA | 2021-2023 | <1%-30% | Largest proposed hydrogen blending program in the U.S.; Led by the National Renewable Energy Laboratory under the Department of Energy, with over 30 industry partners; Aimed at developing a publicly accessible tool to assess blending costs, emission reduction potential and lifecycle impacts [13]. |
| WindGas Falkenhagen | Germany | 2013-2020 | - | Green methane production through wind-powered 2 MW power-to-gas (PtG) facilities [53]. |
| WindGas Hamburg | Germany | 2012-2016 | - | Green hydrogen production using wind power and a 1 MW PEM electrolyzer [54]. |
| Hynet | UK | 2018-now | - | Hydrogen-enabled carbon capture, utilization, and storage (CCUS) is designed to facilitate the decarbonization of a range of industrial sectors, including cement production and waste recycling [55]. |
| PosHYdon | Netherlands | 2019-2024 | - | The first offshore hydrogen project worldwide that produces hydrogen on oil and gas platforms and delivers it through hydrogen-blended pipelines [56]. |
| GRHYD | France | 2014-2019 | 20% | Injected hydrogen into the natural gas distribution network at a 20% volume ratio Demonstrated that hydrogen blending can be safely and effectively implemented without substantial infrastructure upgrades [57]. |
| Sustainable Ameland | Netherlands | 2008 | 20% | Hydrogen blending at concentrations up to 20% caused no adverse effects on end-users, equipment, or pipeline infrastructure [58]. |
| Chaoyang | Liaoning | 2019-2021 | 10% | The first large-scale demonstration project to deliver hydrogen-blended natural gas |
| Renewable Energy Hydrogen Blending Demonstration Project | Province, China | | | in China; Verified the safety and compatibility of existing natural gas pipelines and end-use appliances under actual operating conditions [59]. |
| Yinchuan Ningdong Hydrogen Blending Natural Gas Pipeline Demonstration Platform | Ningxia Hui Autonomous Region, China | **2023** | **24%** | Over a 397 km span, the system completed 100 days of continuous testing, confirming its operational stability and security [60]. |
| Key Technology Research and Application Demonstration of Hydrogen Blending in Natural Gas | Hebei Province, China | **2020-2023** | **5%-20%** | The gas mixture serves commercial, residential, and HCNG vehicle users; Future annual hydrogen delivery to Zhangjiakou is projected to exceed 4 million m³, reducing natural gas use by over 1.5 million m³ and cutting carbon emissions by about 3,000 tons [61]. |
| Research and Pre-commercial Demonstration of Hydrogen Blending, Transportation, and Separation in Natural Gas Pipelines: Zhejiang Energy Pioneer Project | Zhejiang Province, China | **2022-2023** | **30%** | The first integrated demonstration covering urban gas distribution and the full green hydrogen chain in China; From renewable-based production, storage, blending, and transport to gas separation and combustion [62]. |
| Baotou-Linhe Gas Pipeline Project | Nei Mongol Autonomous Region, China | **2023-2024** | **10%** | First long-distance, hydrogen-capable high-pressure pipeline in China; The pipeline extends 249 km, is designed for a pressure of 6.3 MPa, and has a maximum transmission capacity of 1.2 billion m³ per year [63]. |

## III. COORDINATED OPTIMIZATION OF INTERDEPENDENT ELECTRICITY GRID AND HYDROGEN-BLENDED NATURAL GAS NETWORK

The integration of electricity and hydrogen-blended natural gas systems is primarily facilitated through GPP and P2H units. These components serve as critical energy conversion interfaces, introducing complex physical characteristics and operational interactions within the integrated system. To ensure secure, efficient, and economically viable operations, it is essential to develop coordinated optimization strategies. This section reviews recent advancements in modeling, scheduling, planning, and market analysis pertinent to such coordinated operations.

### A. Modeling and simulation of Hydrogen-Penetrated Integrated Gas-Electricity System

Existing modeling techniques for integrated gas-electricity systems (IGES) are well established [64]. However, hydrogen blending introduces new requirements and challenges. First, Hydrogen injection tends to be intermittent and spatially dispersed, resulting in highly nonuniform gas mixtures whose thermophysical properties (density, viscosity) and transport characteristics vary dynamically over time and space. Accurately capturing these transient responses and energy flows requires distributed-parameter models. Second, Future networks may also incorporate diverse unconventional gases such as liquefied natural gas and biogas, rendering traditional calorific value and Wobbe index metrics inadequate for hydrogen-enriched blends [65]. To ensure compatibility with pipeline materials, compressors, and end-use appliances, models must track gas composition and its spatiotemporal distribution rather than relying solely on energy balances [66]. Finally, hydrogen admixture introduces additional governing equations and coupling constraints, including convection dynamics and hydrogen-mass conservation at network nodes, which increase model nonlinearity and nonconvexity. Addressing these complexities requires advanced solution algorithms that enhance computational efficiency and numerical stability [67].

Accurate modeling of hydrogen-blended natural gas networks is crucial for reliable integrated energy system simulations. Ref. [68], [69] utilize three-dimensional computational fluid dynamics to characterize localized hydrogen injection and the resulting nonuniform mixing profiles within pipelines. Ref. [70] uses COMSOL simulations to analyze transient and steady-state behaviors at a node in a looped network composed of one-dimensional pipelines. To reduce computational complexity, several studies introduce simplified pipeline models for CFD-independent simulations. These reduced-order models typically assume steady and uniform flow conditions. For instance, Ref. [71] examines transient wave propagation within a looped network, assuming constant gas composition and fixed gas mass demands. Ref. [72] assumes that injected hydrogen diffuses rapidly throughout the network, employing a uniform transient modeling approach for hydrogen-natural gas mixtures. Ref. [64], [73] focus on converting hydrogen to natural gas with equivalent energy under the same temperature and pressure for calculating energy flow.

The foregoing simplifications prove inadequate for real-world gas networks, which feature multiple sources, non-pipeline elements (e.g., compressors, storage tanks), and intermittent hydrogen injection points [74]. Under such nonuniform and unsteady blending conditions, gas physicochemical properties vary significantly, and traditional models cannot capture the dynamics.

Conventional energy-tracking models, designed for pure natural gas, focus on delivered energy [75], [76]. Consequently, they adhere only to the gas state equation, continuity, momentum, and energy conservation laws, incorporating heating value parameters to represent advective transport of chemical energy [76], [77], as shown in (1).

$$\frac{\partial H_s}{\partial t} + v \frac{\partial H_s}{\partial x} = 0 \qquad (1)$$

where $H_s$ denotes the gas heating value, $t$ represents time, $v$ is the gas velocity, and $x$ is the spatial distance.

However, energy-tracking models only provide a rough estimate of end-user energy efficiency and fails to support detailed management of equipment compatibility, operational safety, and emissions control. Ref. [16], [65], [77] demonstrate that in networks with heterogeneous gas sources, interchangeability cannot be inferred from heating value alone and requires explicit composition tracking.

At the single-pipeline scale, Ref. [77] presents a validated method for tracing chemical-energy flow and mass fractions over an 81.5 km transmission line. Ref. [78] introduces a Lagrangian-coordinate technique that tracks individual gas species and matches its predictions to field data from Polish onshore and Norwegian offshore pipelines. Ref. [74], [77], [78], [79] quantify the impact of hydrogen concentration on compressibility factor, gas constant, hydraulic friction coefficient, and local speed of sound, updating these parameters during simulation. Building on these findings, Ref. [67] analyzes the dynamic behavior of hydrogen-blended networks, identifying a two-stage response and extended transient period that underscore the necessity of dynamic flow models for short-term operations. In these studies, the transport of each gas component is governed by an advection equation.

$$\frac{\partial \phi_k}{\partial t} + v \frac{\partial \phi_k}{\partial x} = 0 \qquad (2)$$

where $\phi_k$ represents the concentration of gas component $k$.

Modeling of intersections poses a common challenge when extending single-pipeline scale model to networks. Ref. [65] addresses this by constructing specialized meshes at intersections to connect all adjacent pipes. Ref. [74] formulates conservation equations for mass and hydrogen fraction transfer, employing Eulerian numerical scheme to avoid the accumulation errors inherent in batch tracking [76], [78] and defining meshes to mitigate numerical instability. Ref. [67] specifies the number of governing equations and boundary conditions to derive the matrix of linearized dynamic flow equations. In these literatures, intersections modeling is conducted as follows: the mass of each component remains constant throughout the mixing process; assuming complete mixing at intersections, the outflows possess homogeneous gas composition.

$$\sum_{N_{in}} M_{in}^i = \sum_{N_{out}} M_{out}^i \qquad (3)$$

$$\sum_{N_{in}} M_{in}^i \phi_{in,k}^i = \sum_{N_{out}} M_{out}^i \phi_{out,k}^i = \phi_{out,k} \sum_{N_{out}} M_{out}^i \qquad (4)$$

where $N_{in}$ and $N_{out}$ denote the number of inflows and outflows at an intersection. $M_{in}^i$ and $M_{out}^i$ are the mass flow rates of the $i$-th inflow and outflow, $\phi_{in,k}^i$ and $\phi_{out,k}^i$ represent the concentration of the $k$-th gas component in the $i$-th inflow and outflow streams.

The aforementioned models, capable of capturing transient gas flow characteristics, are suitable for short-term operations of hydrogen-blended natural gas systems with large time constants. As for systems with short transients or for long-term planning, steady-state models provide a practical alternative by neglecting pipeline dynamic effects and decoupling state variables across time intervals.

At the transmission scale, Ref. [80] develops a steady-state flow model that resolves flow-direction reversals, and validates its results on interconnected electricity and gas networks in Victoria. Ref. [23] analyzes a provincial-scale transmission system in Central China, comparing centralized and distributed hydrogen blending configurations and their impacts on pressure, flow rates, and network loss. Ref. [81] simulates non-isothermal hydrogen injection in a regional transmission network in central Italy, assessing effects on Wobbe index, gas gravity, and high heating value, and determining maximum allowable hydrogen injection of each node.

Within distribution networks, Ref. [82], [83] apply steady flow models to quantify how green hydrogen injection alters nodal pressures and Wobbe indices in low-pressure systems. Ref. [84] investigates urban distribution networks under varying photovoltaic penetration, dynamically tracking gas composition and adapting injection strategies to maximize renewable integration while satisfying demand. Ref. [22] examines the integration of hydrogen with combined-heat-and-power and power-to-hydrogen systems in electricity-heat networks, demonstrating that waste-heat recovery enhances system efficiency and CHP flexibility, reduces operating costs, and identifies the seasonally optimal hydrogen concentration.

TABLE II summarizes the similarities and differences between the steady and transient flow models of conventional natural gas systems [72], [85] and hydrogen-blended natural gas systems [67], [80].

TABLE II Comparison Between Modeling of Traditional Natural Gas And Hydrogen-Blended Natural Gas Systems

| | | Traditional natural gas system | | Hydrogen-blended natural gas system | |
|---|---|---|---|---|---|
| | | steady flow | transient flow | steady flow | transient flow |
| State variables | Mass flow rate | √ | √ | √ | √ |
| | Nodal pressure | √ | √ | √ | √ |
| | Hydrogen concentration | × | × | √ | √ |
| Governing equations and boundary conditions | Flow-pressure Weymouth equation | √ | × | √ | × |
| | Momentum equation and continuity equation | × | √ | × | √ |
| | Mass/energy conservation equation | √ | √ | √ | √ |
| | Pressure of gas supply | √ | √ | √ | √ |
| | Energy consumption of gas loads | √ | √ | √ | √ |
| | Convection equation | × | × | × | √ |
| | Hydrogen mass conservation at intersections | × | × | √ | √ |
| | Composition of gas supply | × | × | √ | √ |
| | Complete mixing assumption | × | × | √ | √ |

Several numerical and analytical methods are available for solving transient gas flow model.

Among numerical approaches, the finite-difference method (FDM) is most widely used [86], [87], [88]. FDM discretizes the domain into a series of nodes and approximates derivatives by Taylor-series expansions, converting partial differential equations into algebraic difference equations. The choice of grid spacing and difference scheme critically influences accuracy and stability, as coarse grids or mismatched time-step sizes can introduce oscillations or nonconvergence. To address this, adaptive step-size control strategies have been developed [89], and a grid-size threshold $\Delta x_{max}$ has been introduced to select stable meshes when tracking varying hydrogen fractions [74].

The finite-volume method (FVM) partitions the domain into control volumes and enforces conservation laws directly. Its inherent stability in convection-dominated flows reduces numerical oscillations, but the stronger mesh dependence and the computational cost of its volume integrals and interpolations limit its use in integrated-energy simulations [78], [90], [91].

The method of characteristics (MOC) transforms first-order partial differential equations into ordinary differential equations along characteristic curves, yielding either analytical solutions or highly accurate approximations [71], [92]. However, it cannot handle higher-order equations, and intersecting characteristics may introduce singularities.

The method of lines (MOL) discretizes spatial derivatives to produce a system of ordinary differential equations in time, which can be solved using standard ODE solvers. This approach accommodates both linear and nonlinear equations, complex boundary conditions, and high-dimensional domains, but its combined spatial and temporal discretization demands substantial computational resources.

Analytical methods seek closed form solutions for system states without discretization or approximation. Functional transformation recasts original differential equations into more tractable forms by introducing new variables or mappings. Common techniques include Laplace transforms [93], Fourier transforms [94], state-space representations [95], [96] and continuous spatial-temporal transformations [97]. However, Fourier and Laplace-based approaches typically neglect initial conditions when solving partial differential equations, limiting their use in period-by-period analysis such as sequential power flow calculation and optimization [98], though they remain suitable for control-oriented studies [75]. An alternative framework analogizes gas networks to electrical circuits to establish a unified energy circuit model of first-order ordinary differential equations, proving effective for energy flow calculations in large scale, heterogeneous systems [99], [100], [101].

For steady flow, both Newton–Raphson and holomorphic embedding (HE) methods are widely employed. The Newton-Raphson algorithm's performance hinges on convergence behavior; for instance, a damping factor introduced in [102] mitigates oscillations and accelerates convergence, while an adaptive damping scheme in [103] dynamically adjusts step sizes to enhance robustness. In contrast, HE is a non-iterative solver [98] that incrementally refines solutions via function approximation. By requiring only a single matrix decomposition, HE offers significant advantages for real-time computation [104]. Moreover, a unified recursive formulation has been proposed to integrate electric, thermal and gas networks, facilitating comprehensive multi-energy flow analysis [105].

## B. Optimal Scheduling of Hydrogen-Blended Integrated Gas-Electricity System

### 1) Optimal scheduling model considering different time scale

Power systems and hydrogen-blended gas networks exhibit disparate inherent time scales, and the dynamics of electricity, gas, and hydrogen subsystems propagate through multi-energy coupling devices. Consequently, HIGES inherently spans multiple temporal resolutions. Its optimal operation is decomposed into three scheduling horizons: long-term (monthly and weekly ahead), day-ahead, and intra-day. Each is governed by distinct influencing factors and resource constraints.

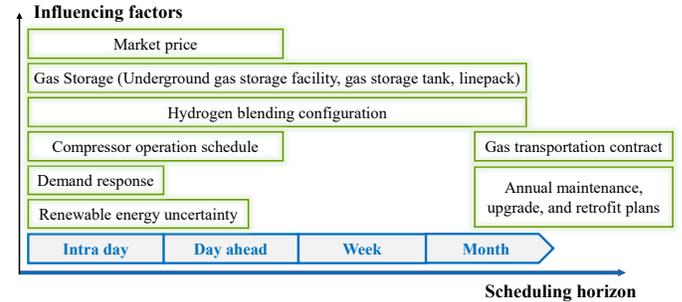

Fig. 2 A summary of different influencing factors for different scheduling horizon

Long-term scheduling incorporates factors such as gas transport contracts, hydrogen-blending configurations, and seasonal storage. Ref. [106] presents a bi-level optimization framework for allocating capacity in hydrogen-blended natural gas pipelines, integrating source selection, transport routing and coordination between the TSO and shippers; The model achieved optimal capacity utilization and equitable cost-sharing under a 5% hydrogen fraction. Ref. [107] develops detailed compressor and pipeline models to minimize energy consumption, achieving an 11.48% reduction in annual operating cost. Ref. [108] optimizes a hydrogen supply chain integrates byproduct hydrogen and natural gas pipelines, reducing the levelized cost of hydrogen by 12.09%. Ref. [109] constructs a dynamic underground salt-cavern storage model and evaluated its emission reduction potential under varying renewable generation and storage capacities. Ref. [109], [110] demonstrates that integrating electrolyzers, seasonal hydrogen storage, and fuel cells enhances medium and long-term renewable energy utilization and reduces reserve requirements. Ref. [111] shows that combining Seasonal storage with hydrogen carriers and power-to-hydrogen-to-power (P2H2P) technology can effectively mitigate seasonal energy imbalances in isolated grids.

Day-ahead scheduling typically relies on next-day forecasts of wind power, solar generation, and electricity demand to determine energy procurement, storage operations (including micro-pumped storage, gas tanks, and hydrogen tanks), and the on/off status of coupling devices such as electrolyzers, fuel cells, and hydrogen-mixing turbines [112]. Intraday scheduling refines these plans using ultra-short-term high-resolution forecasts, which help mitigate forecast errors and enhance overall system efficiency [112]. To address the nonlinear network equations, an extendable algebraic linear approximation (ALA) algorithm has been proposed. Validation over a typical week shows that the location of P2G units significantly influences gas network safety, as reverse gas flows caused by renewable gas injection may pose operational risks. This finding underscores the importance of managing the gas blend even under moderate levels of renewable integration [19]. By integrating high-resolution forecasting, robust day-ahead planning, and rolling optimization via model predictive control (MPC), intraday scheduling achieves both accuracy and supply-demand balance, while accounting for more detailed physical characteristics of the system [113]. Recent advances also include detailed modeling of hydrogen diffusion dynamics [114], optimization of multi-node hydrogen blending considering storage pressure sensitivity [115], and the incorporation of multiphysics processes in hybrid water–biomass electrolysis systems to improve hydrogen production flexibility [116].

### 2) Optimal Scheduling Incorporating System Flexibility

Integrated energy systems (IES) play a critical role in enabling clean, reliable, and economically efficient energy transitions, particularly under high renewable penetration. In HIGES, operational flexibility is supported by energy conversion units, linepack, energy storage components, and demand response. Ref. [117] developed a low-carbon scheduling model that illustrates the benefits of coordinated electricity and hydrogen system operation in managing short-term fluctuations in renewable generation. To further enhance system flexibility, Ref. [118] incorporates the thermal inertia of electrolyzers into day-ahead scheduling models. Ref. [119] quantified gas network adaptability based on pipeline storage, generator capacity and ramping limits. According to Ref. [120], increasing the hydrogen blending ratio can improve wind energy utilization. Ref. [121] observed that hydrogen injection reduces pipeline storage flexibility, while higher hydrogen fraction still yields economic and environmental benefits. Moreover, Ref. [122] introduced a demand response framework that combines price and incentive signals to enhance the flexibility of end-use energy consumption.

### 3) Approaches to Address Uncertainty

Uncertainty in renewable generation, load demand, gas market prices and hydrogen blending ratios significantly complicates scheduling. Two main approaches have been developed to address these uncertainties: stochastic optimization and robust optimization.

Stochastic optimization assumes known probability distributions for uncertain parameters and employs Monte Carlo simulation or analytical transformations to convert the problem into deterministic equivalents, which are typically expectation, multi-scenario or chance-constrained models. For example, historical wind data are fitted to probability density functions and scenarios are generated and reduced via Monte Carlo sampling for day-ahead dispatch in coupled gas–electric systems [123]. A chance-constrained framework for an electricity–HCNG urban integrated energy system handles internal uncertainties from combustion reactions and external fluctuations in load and renewables [124]. However, stochastic methods can be sensitive to scenario choice and may perform poorly out of sample.

Robust optimization dispenses with probabilistic assumptions and instead defines uncertainty sets that bound parameter variations, optimizing against the worst-case realization to ensure feasibility under extreme conditions. A two-stage robust model first optimizes unit commitment in a deterministic setting and then validates feasibility against wind power deviations [125]. Another two-stage formulation for an electricity-heat-gas system with HCNG and power-to-gas units uses a column-and-constraint generation algorithm to manage forecast errors in renewables and demand [126]. To reduce the inherent conservatism of worst-case planning, data-driven uncertainty sets have been proposed; for instance, a minimum-volume enclosing ellipsoid captures correlated wind and solar outputs to generate extreme scenarios [127].

More recently, distributionally robust optimization (DRO) has emerged to combine the strengths of stochastic and robust approaches. DRO constructs ambiguity sets based on partial statistical information such as Kullback–Leibler or Wasserstein distances and minimizes expected cost under the least favorable distribution. A KL-divergence-based DRO model addresses wind uncertainty in a hydrogen-blended energy system [128]. Data-driven DRO using CGAN-generated scenarios and K-medoids clustering has reduced operating cost by 2.1 % compared to classical robust models [129]. A Wasserstein-based chance-constrained dispatch balances cost and risk via a convex conditional value-at-risk formulation [113]. Finally, coordinated scheduling of hydrogen refueling stations and microgrids has been achieved by applying data-driven chance constraints for renewables and load, alongside a Wasserstein-DRO treatment of electricity price volatility [130]. Despite its advantages, DRO entails greater computational complexity, as probability distributions become decision variables.

### 4) Optimal scheduling considering multiple stakeholders

Most existing studies focus on centralized energy management, but HIGES include multiple stakeholders operating across different layers，such as service providers, network operators and end users. Ref. [67] proposes a coordinated dispatch framework that uses dynamic subsidies and nodal gas prices. Its bi-level model maximizes the revenues of both electric and hydrogen-fueled gas operators, improves renewable integration and applies price-based corrective actions when security constraints are violated. For hydrogen blending in transmission pipelines, Ref. [106] develops a bi-level capacity-allocation model in which the network operator determines optimal capacity and tariffs at the upper level, while market participants formulate supply plans based on those tariffs and demand-response strategies, thereby aligning operator and user objectives. Ref. [131] investigates joint optimization among energy producers, system managers, consumers and external suppliers in a regional integrated energy system that includes electric vehicles and adaptive hydrogen blending. Ref. [132] analyses interactions across multiple timescales between coupled electricity–hydrogen–gas systems and distribution networks, applying dynamic game theory to achieve fair pricing.

### 5) Solution algorithm for optimization models

Optimization models for HIGES are typically large-scale, nonlinear, and multi-stage, making global optimization challenging and computationally intensive. Solution methods can be broadly classified into analytical techniques and artificial intelligence (AI) approaches.

Analytical methods include both centralized and decomposition-based strategies. Centralized approaches often apply linearization or convex relaxation to transform complex formulations into linear or mixed-integer programs, which can be solved using standard optimization tools. For instance, Ref. [133] introduces a tightening McCormick algorithm to address bilinear terms in hydrogen mass balance equations. Ref. [121] performs second-order cone relaxation on steady-state gas flow models, followed by penalty functions and boundary-tightening to reduce linearization errors. Ref. [67] decouples convective equations using base-flow approximations and additional physical variables. Decomposition-based methods divide the system into subsystems (e.g., electricity, gas, and hydrogen), which are optimized separately and coordinated through iterative information exchange. Ref. [134] applies the alternating direction method of multipliers (ADMM), enabling parallel computation while preserving data privacy. Ref. [135] develops an enhanced Benders decomposition method for large-scale mixed-integer second-order cone problems, and Ref. [129] adopts a column-and-constraint generation (C&CG) algorithm, which avoids complex reformulations and improves both computational efficiency and cost performance.

To address nonlinear and transient dynamics in pipeline systems more efficiently, AI-based methods have gained attention. Ref. [136] employs neural networks to construct fast surrogate models for near real-time prediction of pipeline pressure and concentration transients. Ref. [137] leverages physics-informed neural networks (PINNs) to solve the partial differential equations governing hydrogen-blended gas networks, effectively integrating physical laws with data-driven learning to enhance simulation accuracy and stability. Ref. [138] applies a multi-objective deep deterministic policy gradient (DDPG) algorithm to jointly optimize operational costs and emissions within a Markov decision process framework. Ref. [139] extends this approach using a multi-agent DDPG algorithm to coordinate interactions among electrolyzers, storage systems, and the power grid. Additionally, population-based algorithms are widely used; for example, Ref. [140] combines Pareto-front analysis with ant colony optimization to address trade-offs among cost, reliability, and environmental performance.

## C. Optimal planning

Optimal planning determines the optimal combination, capacity, location and timing of assets over a given horizon, accounting for representative operating scenarios. In HIGES, planning falls into three categories: energy-station planning, network planning and integrated station-network planning.

Energy-station planning addresses equipment selection and siting for standalone or clustered stations. Ref. [141] develops a multi-objective model for a grid-connected photovoltaic-hydrogen-natural gas facility that emphasizes carbon reduction while coordinating solar generation, hydrogen production and gas usage. Ref. [142] studies capacity sizing and energy management for a hybrid hydrogen-gas turbine with storage, and Ref. [143] evaluates siting and capacity allocation for urban hydrogen refueling stations, considering hydrogen redistribution impacts.

Network planning focuses on the expansion and retrofitting of infrastructure. Ref. [144] examines co-optimization of wind,

hydrogen and electricity networks in the United Kingdom to decarbonize transportation. Ref. [145] proposes a pipeline retrofit model for hydrogen injection that balances capital costs and transport efficiency, while Ref. [146] integrates hydrogen-blended gas pipelines with separation units, selecting the optimal separation technology.

Integrated station-network planning jointly optimizes station and network investments: Ref. [147] presents multi-stage flexible planning of a regional electricity-HCNG system with hydrogen-compatible pipeline upgrades, and Ref. [148] proposes a low-carbon strategy for combined energy systems and hydrogen refueling stations, analyzing the effects of decommissioning conventional fueling stations.

Planning can also be classified by system scale: microgrid, distribution network and transmission network. Microgrid planning applies to small, localized systems, such as communities or industrial parks, which emphasizes self-sufficiency and adaptability [142], [143]. Distribution-level planning covers urban regions and coordinates distributed resources: Ref. [149] minimizes power loss and voltage deviations by optimizing wind turbine placement and soft open points, and Ref. [150] optimizes hydrogen storage configuration in an integrated electricity-gas-heat system with demand response. Transmission-level planning addresses large-scale, long-distance transport: Ref. [108] optimizes a hydrogen supply chain using natural gas pipelines and byproduct hydrogen; Ref. [145], [146] offer models for pipeline retrofitting and hydrogen blending design, respectively. Ref. [151] reviews expansion planning methods for integrated power- gas-hydrogen systems.

Most studies aim to minimize capital and operating costs, though some incorporate carbon emissions, safety and reliability metrics. Ref. [145] demonstrates cost reduction for pipelines and compressors while ensuring transport safety; Ref. [141] prioritizes carbon reduction. Ref. [152], [153] model renewable uncertainty to improve reliability and adaptability. Planning resources include pipeline retrofits [145], [147], hydrogen storage [150], [154], hydrogen pipelines [108], power cables [144], electrolyzers [154], [155], renewable generators [141], [144], fuel cells [144], gas turbines [142] and compressors [146]. Key influencing factors encompass available renewable capacity [144], demand response [156], uncertainty modeling [152], [153], energy market price mechanisms [157] and contingency constraints [158].

## D. Market Exploration

Market coupling between electricity and hydrogen-enriched natural gas reflects both physical dependencies and demand substitution: gas-fired power plants rely on electricity for operation, while end users can shift between power and gas depending on relative availability. Consequently, decisions in one sector exert significant influence on the other. Several studies have examined this interdependence, revealing both economic and security linkages. Ref. [159] observes that gas prices and supply contracts shape unit commitment, economic dispatch and daily scheduling in the power grid, whereas pipeline pressure losses or compressor failures can force thermal units to curtail or shut down. Hence, understanding the synergies in integrated energy systems is crucial. Under constrained information exchange, Ref. [160] develops an equilibrium model for joint electricity–gas operations, demonstrating that the temporal and spatial resolution of shared data materially affects cross-system pricing and cost allocation. Improved data exchange, the study shows, enhances overall system efficiency. Ref. [161] integrates Monte Carlo-based failure scenarios into a mid-term scheduling model derived from day-ahead market clearing, quantifying impacts on electricity prices, unit commitments, dispatch and the generation mix, and identifying optimal locations for additional gas storage to mitigate disruptions. Ref. [162] formulates a bilateral market equilibrium under competitive conditions, coupling renewable generation uncertainty with gas demand in a unified clearing process, and quantifying price distortions and the value of flexibility. Ref. [163] proposes a two-stage stochastic market model that concurrently addresses day-ahead and real-time scheduling, highlighting how pipeline flexibility influences capacity valuation and pricing under high renewable penetration. Ref. [164] analyzes how pipeline constraints and gas-fired generation levels affect the linkage between electricity and gas markets. From an investment perspective, Ref. [165] shows that power-to-gas deployment can flatten electricity price curves, underscoring its significance in long-term asset allocation.

Advances in hydrogen technology and supportive policies have further strengthened market coupling. In Ref. [166], a European case study demonstrates that renewable hydrogen mandates drive power-to-gas capacity expansion and alter electricity and gas price distributions, yielding notable redistributive effects on social welfare. Ref. [132] employs a multi-timescale dynamic game to explore strategic interactions and price-signal formation in distribution-level electricity–hydrogen–gas systems, revealing complex market responses once hydrogen participates. Ref. [167] applies a partial equilibrium framework to assess power-to-gas as a flexibility option in a short-term joint electricity-hydrogen market, quantifying the interplay between demand response and price signals. These studies collectively examine a range of aspects, including market design, price formation, and investment incentives. They provide a comprehensive mapping of the multi-scale interactions and optimization pathways within electrified, hydrogen-integrated energy systems. This detailed analysis offers valuable insights that are crucial for the development of energy markets characterized by efficiency, flexibility, and equity.

Another research aspect examines how to maximize stakeholder benefits in integrated energy markets by refining trading frameworks, game-theoretic mechanisms and revenue-allocation methods to enhance both profitability and equity. In cooperative settings, Ref. [168] develops a hydrogen-natural gas hybrid storage model under a carbon-neutrality mandate, applying coalition-based revenue allocation to achieve simultaneous carbon reduction and economic gain. Ref. [169] coordinates hydrogen blending and power dispatch within a virtual power plant and load aggregator framework, using a distributed algorithm to preserve participant privacy. Ref. [170] adopts a Shapley value-based allocation mechanism to distribute benefits equitably among coalition members. In noncooperative and market-bidding contexts, Ref. [171] proposes a peer-to-peer electricity-hydrogen trading model that integrates hydrogen transport time into pricing strategies, thereby improving transaction efficiency. Ref. [172] formulates a Cournot equilibrium to assess how oligopolistic producers' output and pricing decisions in hydrogen and electricity markets affect overall system profitability. Ref. [173], using a Stackelberg framework in which system operators lead and users and suppliers follow, employs bi-level optimization to maximize profits in an electricity-heat-hydrogen dispatch and pricing scheme. Ref. [174] further develops a competitive planning model for liberalized markets, exploring investment behavior in electricity, gas, hydrogen production and vector-coupling storage under oligopolistic conditions.

Several studies also focus on pricing mechanisms in coupled energy systems. Ref. [175] introduces a multi-energy pricing approach that accounts for time delays in energy transfer and examines price correlations across carriers. Ref. [176], [177] present decomposition methods for hydrogen nodal pricing, dividing prices into components including energy value, carbon reduction and congestion costs, and offer numerical solution techniques and case studies to validate their frameworks. Ref. [178] constructs a dynamic energy–carbon flow model for cost allocation, accurately assigning carbon emission responsibilities and energy charges to each user node to improve transparency and fairness. Ref. [179] employs a joint market-clearing model to derive nodal energy prices and quantify the value of hydrogen blending for decarbonization. Ref. [180] develops a nodal pricing model for gas, hydrogen and their mixtures in mixed-transport pipelines, deriving prices via Lagrangian duality and introducing decarbonization premium to quantify consumers' carbon reduction benefits. Finally, Ref. [181] integrates a hydrogen supply chain model with power dispatch to evaluate how spatially differentiated electricity prices deliver economic signals, demonstrating that location-based pricing can reduce hydrogen delivery costs and network congestion charges, thereby guiding investment decisions in hydrogen infrastructure. Collectively, these studies provide a theoretical foundation and practical pathways for pricing schemes that respect physical flow constraints while incentivizing low-carbon transformation.

## IV. COORDINATED SECURITY MECHANISM AND CONTROL STRATEGY

Owing to the integration of multiple energy carriers, fault propagation of HIGES has become more complex, elevating the risk of cascading failures across subsystems. Studies indicate that energy hubs and coupling devices are key pathways for fault transmission, where imbalances in one subsystem can quickly trigger chain reactions in others, posing serious threats to overall system security and reliability [182], [183]. In integrated electricity-gas systems (IEGS), various interactions may amplify local disturbances into system-wide disruptions [184]. For example, inadequate gas supply to gas turbines, particularly under interruptible contracts, can result in significant power capacity loss. Conversely, rapid ramping of gas-fired units or power outages at compressor stations can reduce gas transport capacity, leading to unsafe pressure drops at pipeline endpoints.

Several risky scenarios are as follows. High penetration of renewable energy increases the frequency of dispatch for gas-fired plants, causing large pressure fluctuations in gas pipelines and undermining transport reliability [20]. Similarly, extreme cold weather can lead to concurrent surges in electricity and gas demand, overloading compressors and lowering the inlet pressure at gas turbines. Moreover, the increased electricity consumption by compressors further intensifies the burden on the power system [185], [186], [187], [188].

Several major incidents have highlighted the vulnerability of multi-energy systems to such interdependencies.

- In February 2021, a severe cold wave in Texas caused widespread freeze-offs at gas wellheads and blockages in pipelines, reducing gas production by about 45% [189]. As a result, gas-fired generation lost up to 26.5 GW, and wind farms lost up to 18 GW due to icing. The surge in electricity demand led to a frequency drop to 59.302 Hz, prompting ERCOT to shed 20 GW of load. The blackout affected 4.8 million users, with some outages lasting up to 70 hours [190], [191].

- In August 2019, a lightning strike in the UK caused transmission lines and distributed generators to disconnect. This event led to the tripping of both steam and gas units at the Little Barford power station, resulting in a loss of 1,878 MW, which accounted for 6.5% of the total load. The power loss caused the frequency to drop to 48.8 Hz, triggering under-frequency load shedding and widespread blackouts [192], [193].

- In August 2017, operator error in Taiwan's gas system caused six gas-fired units at the Datan power plant to trip, resulting in a 4,384 MW capacity loss and affecting 6.68 million users [194].

- In October 2015, a major gas leak at the Aliso Canyon storage facility in California released about 100,000 tons of natural gas [195]. The resulting supply shortage reduced gas-fired generation in Southern California by 2,100 MW during peak summer demand. A technical report later noted that rapid ramping of gas plants could outpace supply response, leading to unsafe inlet pressure and generator trips [196].

- In February 2011, extreme cold weather across the U.S.–Mexico border disrupted gas production, processing, and transport, significantly reducing output in western Texas, New Mexico, and northern Mexico. Interruptible gas contracts were curtailed, and concurrent power outages disabled electric-driven compressors, further reducing pipeline pressure. This cross-sector disruption left 1.3 million electricity users without power and interrupted gas service for over 50,000 customers [197], [198].

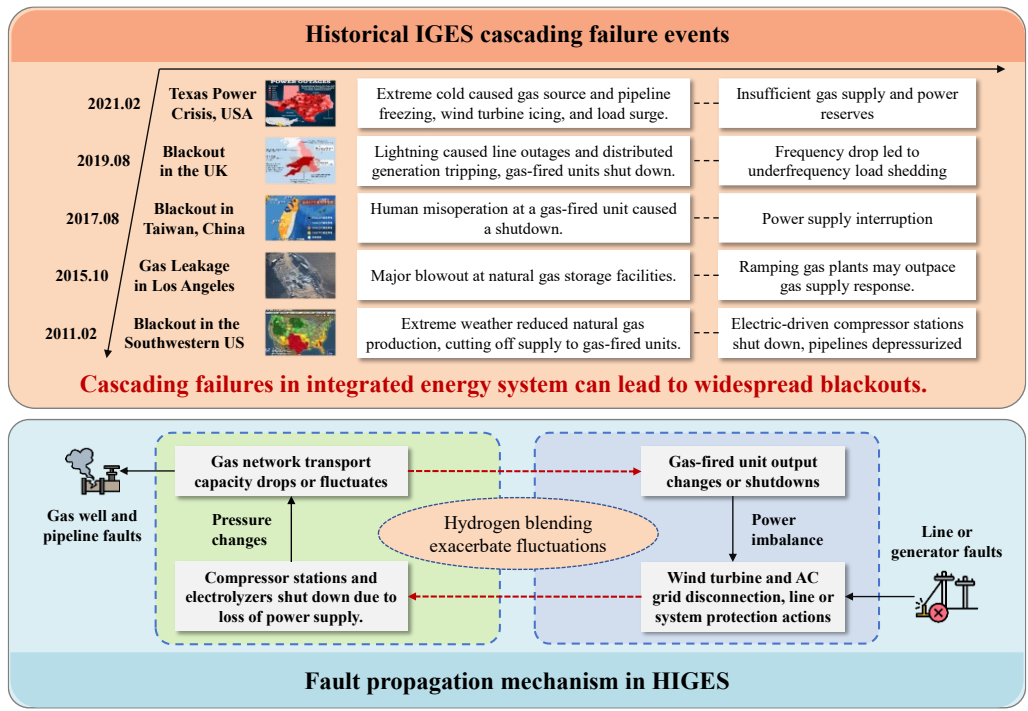

Fig. 3 Historical IGES cascading failure events and the propagation of cascading failures in HIGES

Hydrogen integration in gas-electricity systems introduces unique fault propagation behaviors not seen in conventional configurations. Blending hydrogen changes natural gas properties, accelerating pressure wave propagation and amplifying disturbance magnitudes, which intensifies impacts on downstream nodes [199]. It also reduces pipeline capacity, risking supply shortfalls for critical loads [14]. The high diffusivity of hydrogen and its tendency to embrittle steel accelerate the growth of microdefects into leaks or ruptures under lower pressures and shorter timescales [46], [200]. Furthermore, hydrogen blending affects equipment performance, including reduced metering accuracy, decreased pressure control precision, altered compressor operating envelopes, and combustion instability in gas turbines [201]. Equipment performance is likewise affected: metering accuracy declines, pressure control becomes less precise, compressor operating envelopes shift, and gas turbine combustion may become unstable [201]. Despite these challenges, fault propagation under hydrogen-blended conditions remains largely unstudied. Therefore, existing electricity-gas cascading failure models must be extended to incorporate hydrogen-related characteristics to improve system security and resilience.

### A. Safety Mechanism

Although no incidents have yet been recorded in HIGES, lessons from traditional IGES offer valuable insights into potential safety mechanisms. Existing studies focus on characterizing how disturbances and faults propagate and impact coupled infrastructures. For instance, Ref. [202] simulates the transient response of long pipelines to ruptures, revealing strikingly different pressure and flow behaviors between high- and low-pressure segments. Ref. [203] proposes a life-cycle risk model for hydrogen-blended pipelines under normal operating conditions, assessing hazards from hydrogen-induced cracking, crack propagation, and eventual perforation. Ref. [204] presents a Markov-chain framework to quantify pipeline failure probabilities and the ensuing leak propagation through integrated gas–electric systems. Ref. [205] investigates seismic events, quantifying pipeline leak rates, transmission-line outages, and the resulting load curtailment caused by limited generation capacity and line overloads. Collectively, these methodologies establish a robust foundation for rigorous safety assessment and risk mitigation in future hydrogen-integrated energy networks.

Some studies have quantified these interactions more rigorously. Ref. [206] derives closed-form sensitivity expressions that link gas-network state variables to control inputs, avoiding the opacity of Jacobian-based methods and elucidating how adding storage tanks or adjusting pipeline parameters alters coupled dynamics. Ref. [207] develops a static pressure-to-power-injection sensitivity matrix within a unified power flow framework to pinpoint the most vulnerable gas nodes under blended operation. Using a simplified pipeline model, Ref. [188] assesses short-term power-system security by quantifying how gas-infrastructure outages affect locational marginal prices and post-contingency power flows. Building on this foundation, Ref. [208] incorporates a detailed transient gas-transmission model that more accurately captures the dynamic interplay between gas transport and electricity dispatch.

While previous studies primarily focus on unidirectional fault impacts between subsystems, recent research has advanced the understanding of bidirectional cascading failures in IGES. Complex network theory [209], [210], [211] and co-simulation techniques [212], [213], [214] have been employed to investigate explore fault evolution paths [215]. For instance, a topology-probability model utilizing self-organizing mapping neural networks has been developed to identify vulnerable nodes within the network [209]. Cascading failure simulations combined with random forest algorithms enable rapid vulnerability predictions under various operating conditions [210]. Structural robustness of gas–electric systems has been quantified using static and dynamic indicators derived from complex network metrics [211]. Integrated simulation approaches have been proposed to model cascading failures, particularly focusing on the operational mode transitions of energy-coupling components such as electric-driven compressors [213]. Evaluation metrics for cascading failure consequences have been established to assess the severity and impact of such events [212]. Furthermore, a data fusion method based on energy circuit theory has been introduced to enhance dynamic flow simulations, achieving computational efficiency improvements by over 10 times compared to traditional finite difference methods [214]. Comprehensive reviews have summarized the latest advancements in cascading failure analysis of gas–electric systems, highlighting key modeling, simulation, and risk assessment techniques, and identifying gaps in current research, particularly concerning system vulnerabilities under multi-energy coupling scenarios [215].

### B. Safety Analysis

#### 1) State Estimation

Accurate system boundary conditions are indispensable for security evaluation. State estimation bridges the gap between accessible measurement data and internal system states. State estimation for power systems is now quite mature [216]. Ref. [217], [218], [219] suggest some Kalman-filter-based dynamic state estimation methods for gas networks. However, their iterative solution format makes it hard to incorporate bad data identification, and the analysis targets only simple topologies. Based on unified energy theory, Ref. [220] further presents a dynamic state estimation model for natural gas systems, whose effectiveness in filtering measurement noise, identifying bad data, and supplementing missing measurements is verified. To solve the asynchronization of measurement information and state-execution cycles among different energy systems in IGES, Ref. [221] presents an asynchronous fusion state estimation framework. It combines fixed-period dynamic estimation of natural gas systems, static estimation of power systems, and boundary information interaction between the two. Thus, the consistency of multi-energy system boundary state quantities is guaranteed.

#### 2) Safety assessment and early warning

The purpose of security assessment is to evaluate the operating state of integrated system, identify potential hazards and their severity, issue early warnings, pinpoint vulnerabilities, and guide corrective actions. Security assessment methods are summarized in [222], including point-by-point verification and region-boundary distance methods. Power systems often verify fault resilience through N-1 contingency analysis, which mainly involves defining a contingency set, conducting rapid scans, and performing detailed analyses [223], [224]. Ref. [225] extends the N-1 static security analysis method from power systems to integrated energy systems (IES). Ref. [226] analyzes the impact of N-1 faults in natural gas systems on power systems based on the steady-state models of both systems. The concept of power system security region [227] has also been generalized to IES [228], with preliminary research carried out on static security regions [229] and dynamic security regions [230]. Ref. [231] proposes a robust security region model for wind-power-gas systems, ensuring operational safety within the security region under any wind power injection condition. However, the above research still lacks consideration of the impact of hydrogen, such as the constraints on gas quality, which may introduce additional dimensions to the security region.

### 3) Reliability Analysis and Enhancement

Ref. [16], [232], [233] establish a comprehensive reliability assessment framework for HIGES across short-term, medium-term, and long-term horizons. In Ref. [232], a universal generating function method is employed for short-term reliability evaluation, introducing novel indices to quantify gas adequacy and interchangeability under uncertainty. Ref. [233] applies sequential Monte Carlo simulation over 8 760 hours to capture the temporal feature of renewable energy. The effects of the P2HM process and the fractions and physical properties of hydrogen on the reliability of the IEGSs are analyzed. Ref. [16] a multi-state reliability model for pipelines to characterize long-term corrosion evolution and hydrogen embrittlement. Ref. [234] establishes reliability multi-state models of IGES considering renewable energy uncertainty and cascading effects, showing that carbon emission costs reduce system reliability by affecting wind power penetration. A dynamic cascading-failure analysis has been incorporated into nodal reliability evaluation to illuminate the spatiotemporal propagation of faults [235]. Finally, Ref. [236] evaluates the impact of power-to-gas and combined-heat-and-power unit sizing and siting on the joint reliability of electricity and gas networks. Together, these studies provide a layered understanding of reliability challenges and guide the development of resilient hydrogen-blended energy infrastructures.

### 4) Flexibility Analysis and Enhancement

The flexibility resources in HIGES encompass not only dispatchable generators and demand-response loads typical of power systems but also non-electric assets such as coupling equipment, natural gas networks, and gas-consuming loads. By leveraging these heterogeneous flexibility resources and enabling cross-vector coordination, HIGES can more effectively absorb the variability and uncertainty of wind and solar generation, thereby supporting economically efficient and reliable system operation.

Research on flexibility analysis and enhancement of HIGES has been explored from multiple perspectives. At the equipment level, studies have found that multiple power-to-gas units can aggregate flexibility. These units convert excess electricity into storable gas and hydrogen, helping to balance supply and demand fluctuations and mitigate the impact of renewable energy intermittency on the system [237]. Hydrogen production systems can also adjust their output based on renewable energy availability, offering flexibility to support large-scale renewable integration [238].

In terms of system optimization, an optimal dispatch method for integrated electricity and gas systems has been proposed. This method considers hydrogen injection as a flexible resource to optimize resource allocation and reduce costs. It enables the system to adjust gas network pressure and flow through hydrogen injection, achieving optimal operation of the integrated system [239]. Additionally, research has explored flexibility operations for integrated energy systems with hydrogen under inertia characteristics and carbon trading mechanisms [240]. Distributionally robust flexibility planning has also been developed to address uncertainties, ensuring system flexibility and reliability under different scenarios [241].

At the market mechanism level, the value of power-to-gas as a flexibility option in integrated electricity and hydrogen markets has been examined. It can participate in markets as a flexible resource, providing ancillary services and improving market efficiency [167]. Furthermore, the aggregate power flexibility of multi-energy systems supported by dynamic networks has been analyzed. Dynamic networks facilitate the interaction and transaction of multi-energy systems, enabling better utilization of flexibility through market mechanisms [242].

### 5) Resilience Assessment and Enhancement

Resilience denotes the ability to adjust system operations to minimize losses during major disturbances and to restore normal function swiftly. Ref. [243] employs dynamic gas-composition tracking to assess how hydrogen blending influences gas interchangeability resilience. Ref. [244] optimizes the timing of hydrogen and methane injection, demonstrating through simulation that a resilience-oriented HCNG operation harnesses wind power more effectively to reduce load shedding during energy crises than conventional hydrogen-only injection. Resilience assessment techniques include Monte Carlo simulation and analytical state enumeration [245]. Ref. [246], [247], [248] develop integrated resilience models under windstorm and seismic scenarios. Ref. [249] incorporates gas-thermal inertia as a reserve resource, quantifying its role in enhancing system flexibility during degradation and recovery phases. To strengthen resilience, research has focused on robust optimization and scenario-based approaches [245]. Ref. [250], [251] introduce a tri-level defender-attacker-defender framework to bolster worst-case resilience, while Ref. [252] proposes a network-hardening model to determine optimal reinforcement of pipelines and transmission lines. A scenario-based method in Ref. [253] identifies system vulnerabilities using historical disaster data, albeit with high computational cost. Ref. [254] addresses resilience enhancement from an integrated energy-system planning perspective, and Ref. [245] further presents a unified framework for multi-disaster resilience planning.

## C. Coordinated Control strategies

Control strategies for cascading failures in hybrid integrated gas-electricity systems cover three stages: pre-fault prevention and correction, in-fault intervention, and post-fault recovery. By reviewing key studies, it highlights methods to enhance system stability and reliability.

Pre-fault strategies aim to avert faults before they occur. Ref. [255] introduces a security-constrained preventive control scheme for gas-electric networks. Ref. [256] employs a decision-tree approach on historical data to generate control policies rapidly. Ref. [257] balances control cost against action count to eliminate inefficient interventions. Ref. [258] reduces the complexity of security‐constrained optimal power flow models through iterative binding‐contingency identification, while Ref. [259] replaces finite differences with space-time orthogonal collocation for more accurate and stable corrective control. Ref. [260] integrates these measures into a unified preventive-corrective framework. Preventive control does not consider readjustment of strategies following a fault, making it safer and more conservative but often comes at a higher cost. In contrast, corrective control allows for adjustments after a fault occurs, resulting in a more flexible control strategy. In-fault control seeks to contain fault propagation. Conventional measures include generator tripping, load shedding, network reconfiguration, and FACTS devices [261]. Ref. [262] incorporates a dynamic gas-pressure regulator model into emergency controls for integrated systems. Ref. [263] develops an early-warning scheme that proactively reduces turbine loading to mitigate outages from pipeline failures. Post-fault recovery focuses on rapid system restoration. Ref. [264] proposes a coupled electricity-gas restoration strategy using distributionally robust, chance-constrained programming to balance load recovery with economic efficiency, demonstrating swift network reconstruction after faults.Preventive control offers robust safety at higher cost, while corrective measures provide flexible, real-time adjustments. Together, these approaches form a comprehensive toolkit for managing cascading failures in hybrid gas-electric systems.

## V. SUGGESTIONS ON FUTURE RESEARCH

### A. AI-Enhanced HIGES Modeling and Simulation Techniques

Hybrid hydrogen-natural gas-electricity systems (HIGES), given their multitude of heterogeneous components, intricate energy conversion processes, and exposure to significant environmental uncertainties during practical operation, there is a pressing need for highly efficient and precise modeling and simulation tools. Data-driven techniques offer a promising avenue for addressing traditional physical problems. Existing research has already explored the use of AI for system modeling [265], energy management [266], fault monitoring [267], and reliability evaluation [268]. By integrating AI with HIGES modeling and simulation, it is possible to better meet the demands for real-time simulation and decision-making support in large-scale and complex scenarios, thereby providing robust technical support for the safe and economical operation of integrated energy systems.

### B. HIGES Planning and Scheduling Optimization Under Decentralized Hydrogen Blending Mode

As hydrogen demand grows, centralized blending stations alone cannot satisfy transmission-network-scale energy requirements. Distributed hydrogen-blended gas stations must therefore be sited and sized according to geographic, resource, and safety constraints. Moreover, large-scale hydrogen transport calls for parallel investment in dedicated hydrogen pipelines alongside retrofitting existing gas networks for hydrogen compatibility [269].

Optimal operation of distributed hydrogen blending HIGES requires further research in three key areas: multi-energy integration, flexibility and ancillary services, and demand response and customer engagement. Multi-energy integration involves coordinated optimization of electricity, gas, and hydrogen infrastructures to enhance overall system efficiency, reliability, and economic performance through improved asset sizing and dispatch strategies. Flexibility and ancillary services focus on exploiting the adjustable operation of components such as electrolyzers and gas turbines to support frequency regulation, voltage control, and reserve provision, ensuring system stability under renewable generation variability. Demand response and customer engagement emphasize the active participation of end-users by encouraging hydrogen production during off-peak electricity periods, such as incentivizing hydrogen fuel cell vehicle charging at night, to balance supply and demand, reduce operational costs, and improve system-wide performance.

### C. Market Mechanism Design for HIGES

The market design for HIGES is more complex than that of traditional electricity or natural gas systems due to the involvement of multiple stakeholders across stages such as hydrogen production, blending, transportation, and separation, as well as participation in various markets including electricity, gas, hydrogen, carbon, and flexibility markets. Future research should focus on developing coordinated mechanisms that support multi-stakeholder and multi-market integration. Priorities include designing fair and efficient trading rules to align interests and foster collaboration, establishing pricing mechanisms that reflect supply-demand and cost dynamics, and exploring market coupling strategies to link energy, carbon, and flexibility markets.

### D. Early Warning and Preventive Control Technologies for HIGES

In practical multi-energy system operations, subsystems exhibit functional independence with limited information exchange, impeding effective identification of coupled failures and cascading effects. Consequently, future research should adopt multidimensional methodologies to establish systematic, dynamic fault identification and proactive control mechanisms. First, mathematical approaches should investigate energy-coupling characteristics and dynamic fault propagation between power systems and hydrogen-blended natural gas networks by developing equivalent fault models at critical electro-gas interfaces and dynamic energy flow frameworks to theoretically elucidate intrinsic failure pathways. Second, physical simulations must validate fault evolution patterns under multi-field coupling through transient response testing of critical equipment during faults and dynamic behavior analysis of hydrogen networks. Finally, data-driven analytics utilizing datasets from these mathematical and physical models should determine trigger conditions and root causation of cascading failures in coupled energy systems, enabling dynamic energy flow-based fault path prediction to initiate early warnings and implement targeted preventive measures during incipient failure stages, thereby enhancing safety and operational reliability in HIGES.

## VI. CONCLUSION

This paper reviews the key technologies, modeling methods, coordinated optimization, and safety assessments of HIGES, based on recent research and demonstration projects. The hydrogen-blended natural gas supply chain is well-established and has ongoing demonstration projects, making it a promising low-cost solution for transitioning to a future electricity-hydrogen system. In terms of modeling, steady-state methods are well developed. However, dynamic modeling of fluid networks that captures distributed parameters and transient behaviors still faces challenges in terms of accuracy, scalability, and computational efficiency. Most optimization studies focus on specific scenarios. Future research should explore leveraging the complementary characteristics of heterogeneous energy sources under new safety constraints to achieve coordinated optimization across economic, low-carbon, and reliability objectives. Efficient solutions for complex nonlinear mixed-integer problems are also challenging. Market mechanisms are still in the early stages and require further exploration to develop fair trading frameworks. This would enable optimal resource allocation across electricity, gas, hydrogen, and carbon markets. Safety and control methods from integrated energy systems can be adapted to HIGES, but studies on how hydrogen blending affects the multi-timescale dynamic responses of coupled systems are still lacking. The research gaps identified in this paper offer valuable insights for both academic research and engineering applications.